\documentclass[preprint,tightenlines,showpacs,showkeys,floatfix,nofootinbib,
superscriptaddress,fleqn]{revtex4}
\usepackage{graphicx}
\usepackage{bm}
\usepackage{dcolumn}
\usepackage{amsmath}
\usepackage{amsfonts}
\usepackage{amssymb}

\begin{document}

\preprint{PNU-NTG-13/2005}
\title{Electric properties of the baryon anti-decuplet in the
SU(3) chiral quark-soliton model}
\author{T. Ledwig}
\email{Tim.Ledwig@tp2.rub.de}
\affiliation{Institut f\"ur Theoretische Physik II, Ruhr-Universit\" at Bochum,
D--44780 Bochum, Germany}
\author{H.-Ch. Kim}
\email{hchkim@pusan.ac.kr}
\affiliation{Department of Physics, and Nuclear Physics \&
Radiation Technology Institute (NuRI), Pusan National University, 609-735
Busan, Republic of Korea}
\author{A. J. Silva}
\email{ajose@teor.fis.uc.pt}
\affiliation{Departamento de F\'\i sica and Centro de F\'\i sica Computacional,
Universidade de Coimbra, P-3000 Coimbra, Portugal}
\affiliation{Faculdade de Engenharia da
Universidade do Porto, R. Dr. Roberto Frias s/n, P-4200-465 Porto,
Portugal}
\author{K. Goeke}
\email{Klaus.Goeke@tp2.rub.de}
\affiliation{Institut f\"ur Theoretische Physik II, Ruhr-Universit\" at Bochum,
D--44780 Bochum, Germany}
\date{March 2006}

\begin{abstract}
We investigate the electric form factors and charge radii of the
pentaquark baryons within the framework of the chiral
quark-soliton model.   We consider the rotational $1/N_c$ and
linear $m_s$ corrections, assuming isospin symmetry and employing
the symmetry-conserving quantization.  The flavor-decomposed
charge densities of the $\Theta^+$ are presented.  The electric
form factors and charge radii of the charged pentaquark baryons
turn out to be very similar to those of the corresponding octet
baryons.  The charge radii of the neutral pentaquark baryons are
obtained to be very tiny and positive.  The strange electric form
factor of the pentaquark proton is shown to be larger than the
corresponding one of the proton by around $20\, \%$.  We also
present the charge radii of the baryon decuplet for comparison.

\end{abstract}
 \maketitle

%\section{Introduction}
{\bf 1.} The pentaquark baryons have been a hot issue, since
Diakonov {\it et al.} predicted the mass and width of the
pentaquark baryon $\Theta^+$~\cite{Diakonov:1997mm} with strange
quark number $S=+1$ and valence quark structure $uudd\bar{s}$. An
earlier estimate of the mass in the soliton approach of the Skyrme
model was given by Praszalowicz~\cite{Praszalowicz}. A great deal
of experimental and theoretical works have been already piled up
to confirm its existence and to understand its properties (see,
for example, a recent review~\cite{Hicks:2005gp} for the
experimental results). Many experiments have announced the
existence of the $\Theta^+$ after the first observation by the
LEPS collaboration~\cite{Nakano:2003qx}, while the $\Theta^+$ has
not been seen in almost all high-energy experiments.  Moreover,
exotic $\Xi_{\overline{10}}$ states were observed by the NA49
experiment at CERN~\cite{Alt:2003vb}, though its existence is
still under debate.

A very recent CLAS experiment, a dedicated experiment to search
for the $\Theta^+$, has announced null results of finding the
$\Theta^+$ in $\gamma p \to \bar K^0
\Theta^+$~\cite{Battaglieri:2005er}.  Though this experiment is
high statistics measurement, it is too early to conclude the
absence of the $\Theta^+$, because the previous positive evidences
appeared mostly in the reactions from the neutron and the
kinematic regions of each experiment are different.  Moreover,
several new experiments searching for the $\Theta^+$ are in
progress~\cite{Hotta:2005rh,Miwa:2006if}.  In the present obscure
status for the $\Theta^+$, more efforts should be made for
understanding the $\Theta^+$ theoretically as well as
experimentally.

In addition, a recent GRAAL experiment~\cite{Kuznetsov:2004gy}
announced a new nucleon-like resonance with a seemingly narrow
decay width $\sim 10$ MeV and a mass $\sim 1675$ MeV in the
$\eta$-photoproduction from the neutron target.  This new
nucleon-like resonance, $N^*(1675)$, may be regarded as a
non-strange pentaquark because of its narrow decay width which is
known to be one of the characteristics of typical pentaquark
baryons, though one should not exclude a possibility that it might
be one of the known $\pi N$ resonances (possibly,
$D_{15}$)~\cite{Kuznetsov:2006de}. This GRAAL data is consistent
with the results for the transition magnetic moments in the chiral
quark-soliton model ($\chi$QSM)~\cite{Kim:2005gz} as well as the
partial-wave analysis for the non-strange pentaquark
baryons~\cite{Azimov:2005jj}.  Moreover, a recent theoretical
calculation of the $\gamma N\to \eta N$
reaction~\cite{Choi:2005ki} describes qualitatively well the GRAAL
data, based on the values of the magnetic transition moments in
Refs.~\cite{Kim:2005gz,Azimov:2005jj}, which implies that the
$N^*$ seen in the GRAAL experiment could be favorably identified
as one of the pentaquark baryons.

Being motivated by these new experimental findings, a great deal
of theoretical investigations have been performed, which are
summarized in recent reviews~\cite{Jennings:2003wz,Zhu:2004xa,
Goeke:2004ht}.  Much of these theoretical works has concentrated
on understanding the structure of these exotic pentaquark baryons.
While the mass of the $\Theta^+$ is experimentally known, other
properties such as its spin and parity are undetermined. Moreover,
the electromagetic properties like charge radii and magnetic
moments, and various form factors are experimentally unknown.
Understanding such properties of the pentaquark baryons is,
however, of great importance, since it will shed light on
questions how the pentaquark baryons are constituted in terms of
quarks and how they are produced in different reactions.  In
particular, the electric charge densities of the pentaquark
baryons provide microscopical information on how their valence
quarks and antiquark are located inside them, i.e.,  as already
discussed in Ref.~\cite{Polyakov:2004aq}, electric properties of
the pentaquark baryons reveal their internal quark structure.

Thus, in the present work, we aim at investigating the electric
properties of the baryon anti-decuplet such as electric charge
radii and electric form factors, based on the $\chi$QSM with
isospin symmetry and symmetry-conserving
quantization~\cite{Praszalowicz:1998jm} imposed. The $\chi$QSM has
been proved very successful not only in predicting the $\Theta^+$
but also in describing various properties of SU(3) baryon octet
and decuplet such as the mass splittings and form factors with the
same set of fixed parameters~\cite{Christov:1995vm}.  In
particular, the dependence of almost all form factors on the
momentum transfer is well reproduced within the $\chi$QSM.  As a
result, the parity-violating asymmetries of polarized
electron-proton scattering which require nine different form
factors (six electromagnetic form factors $G_{E,M}^{(u,d,s)}
(Q^2)$ and three axial-vector form factors $G_{A}^{(u,d,s)}
(Q^2)$) are in good agreement with experimental
data~\cite{Silva:2005qm}.  Therefore, it is worthwhile to extend
the study of the form factors to the baryon anti-decuplet within
the $\chi$QSM.  Moreover, as several models suggest, the
pentaquark proton may contain a valence strange quark ($s$) and
antiquark ($\bar{s}$), while the strange quark component of the
proton is excited from the Dirac sea.  Thus, it is also of great
interest to study its strange electric form factor and to compare
it with that of the proton.

The paper is organized as follows: In Section 2, we briefly
review the formalism for the calculation of the electric
properties of the baryon anti-decuplet in the $\chi$QSM.  In
Section 3, we discuss the numerical results for the electric
charge densities of the pentaquark baryons, and for their electric
charge radii and electric form factors.  We also compare the
strange electric form factor of the pentaquark proton with that of
the proton.  In Section 4, we summarize and conclude the present
work.

%\section{Formalism}
{\bf 2.} In this section we briefly review the formalism of the
$\chi$QSM for calculating the electric properties of the baryon
anti-decuplet. We refer to Refs.~\cite{Christov:1995vm,Kim:1995mr}
for details.  The electromagnetic form factors for the baryon
anti-decuplet are expressed in terms of the quark matrix elements:
\begin{equation}
\langle B_{\overline{10}} (p') | J_\mu(0) | B_{\overline{10}}(p)\rangle =
\overline{u}_{B_{\overline{10}}} ({\bm p'},s') \left[\gamma_\mu
  F_1^{B_{\overline{10}}} (Q^2) +   i\frac{\sigma_{\mu\nu}
    q^\nu}{2M_N} F_2^{B_{\overline{10}}} ( Q^2) \right]
u_{B_{\overline{10}}}({\bm p}, s),
\end{equation}
where $Q^2$ is the square of the four momentum transfer $Q^2=-q^2$
with $Q^2 >0$. The $M_N$ and $u_{B_{\overline{10}}}$ denote the
nucleon mass and the Dirac spinor with the corresponding momentum
$\bm p$ and spin $s$ of the baryon anti-decuplet, respectively.
The quark electromagnetic current $J_\mu$ in Euclidean space is
defined in terms of the triplet $(J_\mu^{(3)})$ and octet
$(J_\mu^{(8)})$ currents
\begin{equation}
J_\mu = -i\psi^\dagger \gamma_\mu \hat{Q} \psi = \frac12
\left(J_\mu^{(3)} + \frac1{\sqrt{3}} J_\mu^{(8)} \right)
\end{equation}
with
\begin{equation}
J_\mu^{(3)} = -i\psi^\dagger \gamma_\mu \lambda_3 \psi,\;\;\;
J_\mu^{(8)} = -i\psi^\dagger \gamma_\mu \lambda_8 \psi.
\end{equation}

The Dirac and Pauli form factors can be expressed in terms of the
Sachs electric and magnetic form factors as follows:
\begin{eqnarray}
G_E^{B_{\overline{10}}} ({\bm Q}^2) &=& F_1^{B_{\overline{10}}} ({\bm Q}^2) -
\frac{{\bm Q}^2}{4M_N^{2}} F_2^{B_{\overline{10}}} ({\bm Q}^2) ,\cr
G_M^{B_{\overline{10}}} ({\bm Q}^2) &=& F_1^{B_{\overline{10}}} ({\bm Q}^2) +
F_2^{B_{\overline{10}}} ({\bm Q}^2).
\end{eqnarray}
The electric and magnetic form factors can be  respcetively written as
matrix elements of the time and space components of the
electromagnetic current:
\begin{eqnarray}
\langle B_{\overline{10}} (p') | J_4(0) | B_{\overline{10}}(p)\rangle
&=& G_E^{B_{\overline{10}}} ({\bm Q}^2) \delta_{s's}  \cr
\langle B_{\overline{10}} (p') | J_k(0) | B_{\overline{10}}(p)\rangle
&=& \frac{i}{2M_N} \epsilon_{klm} (\sigma^l)_{s's}  q^m
G_M^{B_{\overline{10}}} ({\bm Q}^2)
\label{eq:ff},
\end{eqnarray}
where $\sigma^l$ stand for the Pauli spin matrices.  Since we are
interested in the electric form factors in this work, we concentrate
on the first equation in Eq.(\ref{eq:ff}).

Having performed a calculation as in Ref.~\cite{Kim:1995mr}, we
arrive at the following expression:
\begin{equation}
G_{E}^{B_{\overline{10}}}({\bm Q}^2)  =
G_{E}^{B_{\overline{10}},m_s^0}({\bm Q}^2) +
G_{E}^{B_{\overline{10}},m_s^1, {\rm op}}({\bm Q}^2) +
G_{E}^{B_{\overline{10}},m_s^1,{\rm wf}}({\bm Q}^2) ,
\end{equation}
where
\begin{eqnarray}
G_{E}^{B_{\overline{10}},m_s^0}({\bm Q}^2) & = & \frac18
\left(\frac{1}{3}\, \mathcal{B}({\bm Q}^2)\,+\,
\,\frac{\mathcal{I}_{1}({\bm Q}^2)}{I_{1}}\,
+\,6\,\frac{\mathcal{I}_{2}({\bm Q}^2)}{I_{2}}\right)
Q_{B_{\overline{10}}},
\label{eq:final1}
\\
G_{E}^{B_{\overline{10}},m_s^1, {\rm op}}({\bm Q}^2) &=&
\frac{m_s}{126I_{1}}\,\left[\begin{array}{c}
18\, Q_{B_{\overline{10}}}\\
11\, Q_{B_{\overline{10}}}+20\\
12\, Q_{B_{\overline{10}}}+32\\
13\, Q_{B_{\overline{10}}}+44\end{array}\right]\,(I_{1}\mathcal{K}_{1}({\bm
Q}^2)-K_{1}\mathcal{I}_{1}({\bm Q}^2))\cr
 &  & + \frac{m_s}{63I_{2}}\,\left[\begin{array}{c}
27\, Q_{B_{\overline{10}}}\\
-\, Q_{B_{\overline{10}}}+23\\
-3\, Q_{B_{\overline{10}}}+20\\
-5\,
Q_{B_{\overline{10}}}+17\end{array}\right]\,(I_{2}\mathcal{K}_{2}({\bm
Q}^2)-K_{2}\mathcal{I}_{2}({\bm Q}^2))\cr
 &  & -m_s \left(\frac{1}{36}\,Q_{B_{\overline{10}}}
-\frac{1}{252}\,\left[\begin{array}{c}
4\, Q_{B_{\overline{10}}}\\
-3\, Q_{B_{\overline{10}}}+6\\
-2\, Q_{B_{\overline{10}}}+4\\
-\, Q_{B_{\overline{10}}}+2\end{array}\right]\,\right)\mathcal{C}({\bm Q}^2),
\label{eq:final2}
\\
G_{E}^{B_{\overline{10}},m_s^1,{\rm wf}}({\bm Q}^2) & = &
\frac{d_8}{6}\,\left[\begin{array}{c}
0\\
-2\, Q_{B_{\overline{10}}}+2\\
-Q_{B_{\overline{10}}}+1\\
0\end{array}\right]\mathcal{B}({\bm
Q}^2)-\frac{d_8}{2}\left[\begin{array}{c}
0\\
-2\, Q_{B_{\overline{10}}}+2\\
-Q_{B_{\overline{10}}}+1\\
0\end{array}\right]\frac{\mathcal{I}_{1}({\bm Q}^2)}{I_{1}}\cr
&-& 2d_{8}\left[\begin{array}{c}
0\\
0\\
0\\
0\end{array}\right]\frac{\mathcal{I}_{2}({\bm Q}^2)}{I_{2}}
+ \frac{d_{27}}{120}\,\left[\begin{array}{c}
0\\
7\, Q_{B_{\overline{10}}}-2\\
6\, Q_{B_{\overline{10}}}+4\\
5\, Q_{B_{\overline{10}}}+10\end{array}\right]\,\mathcal{B}({\bm
Q}^2)\cr
&-&
\frac{7d_{27}}{120}\,\left[\begin{array}{c}
0\\
-7\, Q_{B_{\overline{10}}}+2\\
-6\, Q_{B_{\overline{10}}}-4\\
-5\, Q_{B_{\overline{10}}}-10\end{array}\right]\,
\frac{\mathcal{I}_{1}({\bm Q}^2)}{I_{1}}
- \frac{d_{27}}{36}\,\left[\begin{array}{c}
0\\
21\, Q_{B_{\overline{10}}}-6\\
18\, Q_{B_{\overline{10}}}+12\\
15\, Q_{B_{\overline{10}}}+30\end{array}\right]\,
\frac{\mathcal{I}_{2}({\bm Q}^2) }{I_{2}} \cr
&+& \frac{d_{\overline{35}}}{112}\,\left[\begin{array}{c}
4\, Q_{B_{\overline{10}}}\\
-3\, Q_{B_{\overline{10}}}+6\\
-2\, Q_{B_{\overline{10}}}+4\\
-Q_{B_{\overline{10}}}+2\end{array}\right]\mathcal{B}({\bm Q}^2)
-\frac{d_{\overline{35}}}{112}\,\left[\begin{array}{c}
4\, Q_{B_{\overline{10}}}\\
-3\, Q_{B_{\overline{10}}}+6\\
-2\, Q_{B_{\overline{10}}}+4\\
-1\, Q_{B_{\overline{10}}}+2\end{array}\right]\,
\frac{\mathcal{I}_{1}({\bm Q}^2) }{I_{1}}\cr
&-&
\frac{d_{\overline{35}}}{56}\,\left[\begin{array}{c}
4\, Q_{B_{\overline{10}}}\\
-3\, Q_{B_{\overline{10}}}+6\\
-2\, Q_{B_{\overline{10}}}+4\\
-1\, Q_{B_{\overline{10}}}+2\end{array}\right]\,
\frac{\mathcal{I}_{2}({\bm Q}^2) }{I_{2}}
\label{eq:final3}
\end{eqnarray}
in the basis of $[\Theta^+, N_{\overline{10}},
\Sigma_{\overline{10}}, \Xi_{\overline{10}}]$.  Here, the
coefficients $d_8$, $d_{27}$, and $d_{\overline{35}}$ are
proportional to $m_s$ and known from the SU(3)
algebra~\cite{Yang:2004jr}, the $I_1$ etc. are moments of inertia
whose expressions can be found in Ref~\cite{Christov:1995vm}, and
the other quantities such as ${\cal B} ({\bm Q}^2)$ can be found
in Ref.~\cite{Silva:2002ej}. The $Q_{B_{\overline{10}}}$ denotes
the charge of the corresponding pentaquark baryon.

%\section{Results and discussion}
{\bf 3.} We now discuss the results obtained from the present
work.  A detailed description of the numerical methods can be
found in Ref.~\cite{Kim:1995mr,Christov:1995vm}. We first want to
emphasize that the only free parameter of the $\chi$QSM is the
constituent quark mass $M$.  Since the electric charge radii and
form factors turn out to be insensitive to the constituent quark
mass $M$ in general, we choose in this study the value
$M=420$~MeV, which is found from a best fit to many nucleon
observables~\cite{Christov:1995vm}.  The other parameters of the
model are the current nonstrange quark mass and the cut-off
parameter of the proper-time regularization: They are all fixed
for a given $M$ in such a way that mesonic properties, i.e. the
physical pion mass and decay constant, are well reproduced. The
mass of the strange quark is throughout this work set to $m_{\rm
s}=180$ MeV.

\begin{figure}[h]
\begin{center}
\includegraphics[scale=0.7]{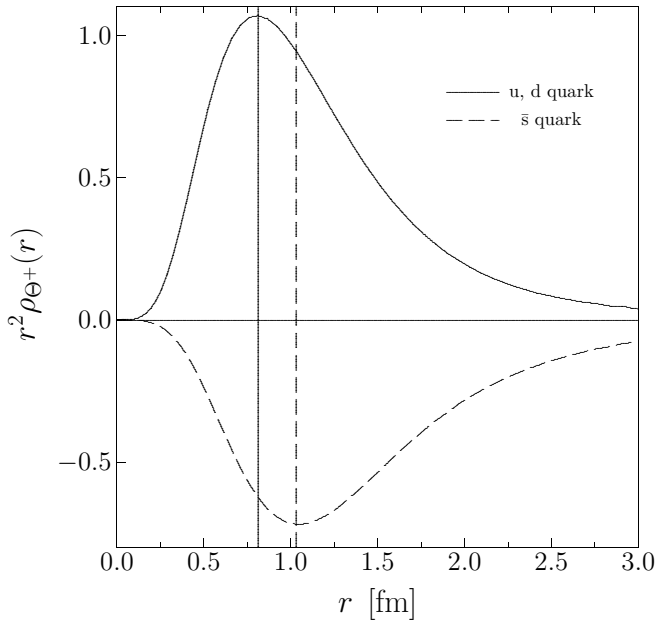}
\end{center}
\caption{Flavor-decomposed electric charge densities of the
pentaquark baryon $\Theta^+$ multiplied by $r^2$ as functions of
$r$.  The solid curves draw both the up ($\rm u$) and down ($\rm
d$) quark components which are identical due to the assumed
isospin symmetry, whereas the dashed curve depicts the strange
($\rm s$) component (multiplied by $(-1)$ for clarity).  The
vertical lines denote the position of the corresponding maximum
values of the densities.} \label{fig:1}
\end{figure}

Figure~\ref{fig:1} depicts the flavor-decomposed electric charge
densities of the pentaquark baryon $\Theta^+$.  The solid curves
draw both the valence up ($\psi^{\dagger}_u \psi_u$) and down
($\psi^{\dagger}_d \psi_d$) quark densities which are identical
due to the assumed isospin symmetry, whereas the dashed curve
represents the strange quark ($\psi^{\dagger}_s \psi_s$) one. The
vertical lines denote the corresponding maximum values of the
densities. These flavor-decomposed densities of the $\Theta^+$
make it possible for us to see the quark configuration inside the
$\Theta^+$.  As shown in Fig.~\ref{fig:1}, the peak of the valence
strange anti-quark is located at slightly larger $r$, compared to
that of the up and down quarks.  It indicates that the strange
anti-quark seems to be most probably found in the outer region of
the $\Theta^+$, though it lies still rather close to the up and
down quarks. Though we present here the densities of the
$\Theta^+$, we find in general that the valence anti-quarks of the
pentaquark baryons tend to be located at larger $r$ and the shapes
of the densities are very similar to those of the $\Theta^+$.

\begin{table}[h]
\begin{tabular}{|c|cccccccccc|}
\hline anti-decuplet& $\Theta^{+}$& $p_{\overline{10}}^{*}$&
$n_{\overline{10}}^{*}$& $\Sigma_{\overline{10}}^{*+}$&
$\Sigma_{\overline{10}}^{*0}$& $\Sigma_{\overline{10}}^{*-}$&
$\Xi_{\overline{10}}^{*+}$& $\Xi_{\overline{10}}^{*0}$&
$\Xi_{\overline{10}}^{*-}$&
$\Xi_{\overline{10}}^{*--}$\tabularnewline \hline $\langle r^{2}
\rangle_{\rm E} [{\rm fm}^{2}]$& $0.770$& $0.771$& $0.014$&
$0.772$& $0.008$& $0.756$& $0.773$& $0.003$& $0.768$&
$0.769$\tabularnewline \hline
\end{tabular}
\caption{The electric charge radii of the baryon anti-decuplet.}
\label{table:1}
\end{table}

\begin{table}[h]
\begin{tabular}{|c|cccccccc|}
\hline octet& $p$& $n$& $\Lambda$& $\Sigma^{+}$& $\Sigma^{0}$&
$\Sigma^{-}$& $\Xi^{0}$& $\Xi^{-}$ \tabularnewline \hline $\langle
r^{2} \rangle_{\rm E} [{\rm fm}^{2}]$& $0.768$& $-0.071$&
$-0.029$& $0.771$& $0.026$& $0.720$& $-0.054$&
$0.707$\tabularnewline \hline
\end{tabular}
\caption{The electric charge radii of the baryon octet.}
\label{table:2}
\end{table}

\begin{table}[h]
\begin{tabular}{|c|cccccccccc|}
\hline decuplet& $\Delta^{++}$& $\Delta^{+}$& $\Delta^{0}$&
$\Delta^{-}$& $\Sigma^{*+}$& $\Sigma^{*0}$& $\Sigma^{*-}$&
$\Xi^{*0}$& $\Xi^{*-}$& $\Omega^{-}$\tabularnewline \hline
$\langle r^{2} \rangle_{\rm E} [{\rm fm}^{2}]$& $0.813$& $0.794$&
$-0.038$& $0.869$& $0.823$& $-0.013$& $0.848$& $0.012$& $0.826$&
$0.805$\tabularnewline \hline
\end{tabular}
\caption{The electric charge radii of the baryon decuplet.}
\label{table:3}
\end{table}

With these densities, we are able to calculate the electric
charge radii $\langle r^{2} \rangle_{\rm E}$ of the baryon
anti-decuplet which are listed in Table~\ref{table:1}. For
comparison, we also list those of the baryon octet and decuplet in
Table~\ref{table:2} and Table~\ref{table:3}, respectively.  It is
interesting to see that the electric charge radius of the
$\Theta^+$ is only slightly larger than that of the octet proton.
 This implies that the $\Theta^+$ pentaquark baryon is quite
compact, even though its main Fock state component corresponds to
five quarks. In addition the SU(3) symmetry-breaking effects are
negligible, as we will see later. We can also find from
Table~\ref{table:1} that the electric charge radii of the baryon
anti-decuplet are in general comparable to those of the baryon
octet in magnitude apart from the neutral baryons.  The electric
charge radii of the neutral baryon anti-decuplet particles turn
out to be much smaller than those of the neutral baryon octet
ones. In particular, the $\langle r^2\rangle_{\rm
E}^{\Xi_{\overline{10}}^{*0}}$ is more than an order of magnitude
smaller than $\langle r^2\rangle_{\rm E}^{\Xi^{0}}$.  The reason
lies in the fact that in the present formalism the electric charge
radii of the neutral baryon anti-decuplet arise only from the SU(3)
symmetry-breaking effects and are thus very tiny, in fact they
vanish when the $m_{\rm s}$ corrections are switched off. We will
discuss it more in detail later in the case of the electric form
factors. The electric charge radii of the baryon decuplet are
found to be larger than those of the octet and anti-decuplet.

\begin{figure}[h]
\begin{center}
\includegraphics[scale=0.7]{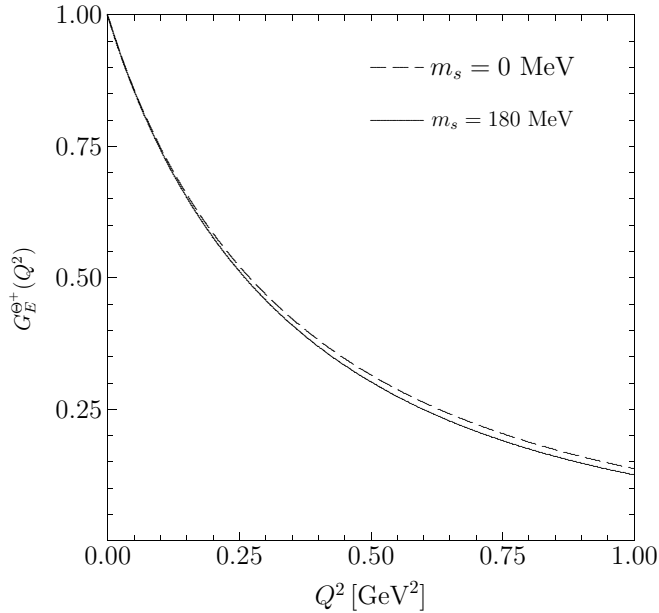}
\end{center}
\caption{Electric form factor of the $\Theta^+$ with and without the
  strange quark mass.  The solid curve depicts the form factor for $m_s=180$ MeV,
  while the dashed one for $m_s =0$.}
\label{fig:2}
\end{figure}

Figure~\ref{fig:2} draws the electric form factor of the
$\Theta^+$ with and without the current strange quark mass $m_{\rm
s}$.  The effect of SU(3) symmetry breaking is almost negligible
for the electric form factor of the $\Theta^+$. This feature holds
basically for all electric form factors of the charged pentaquark
baryons. The reason for this lies in the fact that in the
$\chi$QSM the strange quarks are not valence quarks but arise from
sea quark excitations due to the rotations of the system.

\begin{figure}[h]
\begin{center}
\includegraphics[scale=0.57]{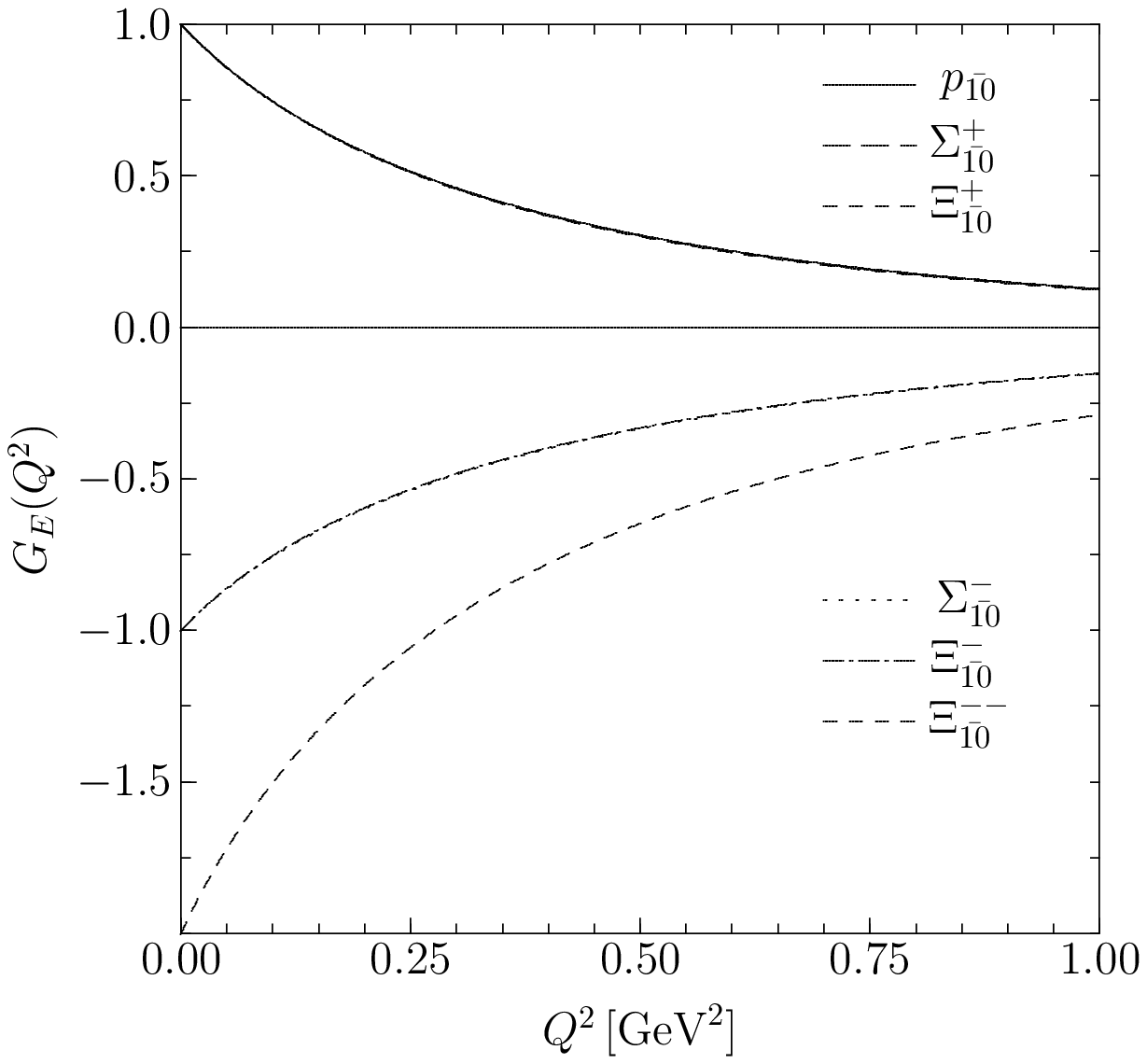} \hspace{0.4cm}
\includegraphics[scale=0.57]{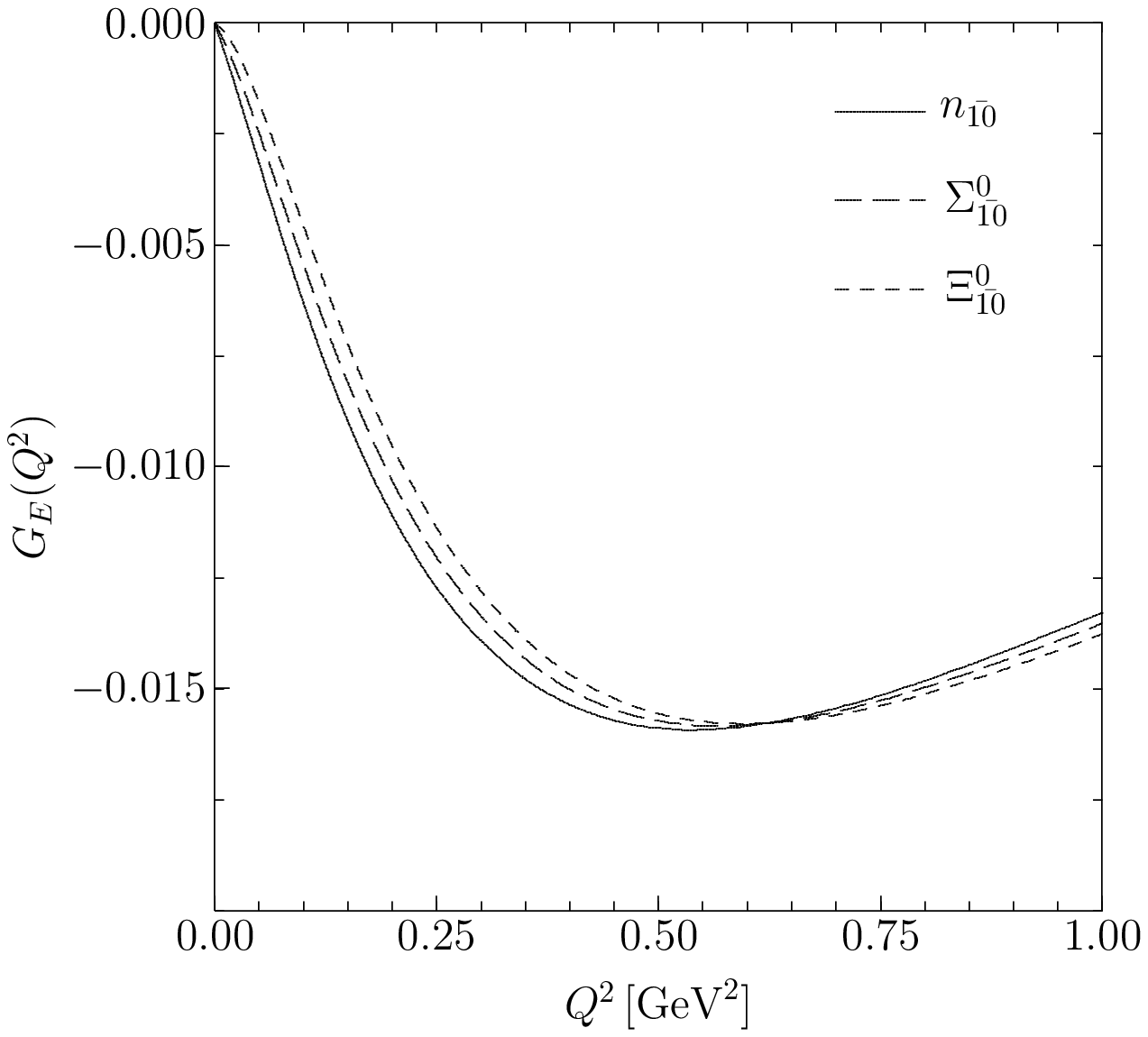}
\end{center}
\caption{Electric form factors of the baryon anti-deucplet as
functions of $Q^2$.  In the left panel we depict those of the
charged pentaquark baryons, while in the right panel we draw those
of the neutral ones.} \label{fig:3}
\end{figure}

In the left panel of Figure~\ref{fig:3} we depict the electric
form factors of the charged baryon anti-decuplet, while in its
right panel we draw those of the neutral baryon anti-decuplet.  It
is interesting to see that there is almost no difference in the
$Q^2$ dependence for charged pentaquark baryons with the same
charge.  It indicates that there is no flavor dependence. The
reason can be found in Eqs.(\ref{eq:final1})-(\ref{eq:final3}).
The leading and rotational $1/N_c$ terms of the whole
anti-decuplet are expressed in the first term of
Eq.(\ref{eq:final1}), which is just proportional to the charge of
the corresponding pentaquark baryon $Q_{B_{\overline{10}}}$. It
means that the electric form factors of the pentaquark baryons
with the same charge are identical when the SU(3) symmetry
breaking is turned off. However, the effects of the SU(3) symmetry
breaking on the electric form factors are in general small.

Moreover, the various terms of the $m_{\rm s}$ corrections are
destructively interfering for the electric form factors of the
baryon anti-decuplet as shown in Eqs.(\ref{eq:final2}) and
(\ref{eq:final3}), so that the symmetry-breaking terms turn out to
be negligibly small. As a result, the electric form factors of the
charged pentaquark baryons appear to be almost the same.

In pure SU(3) symmetry, the electric form factors of the neutral
pentaquark baryons vanish, since they are solely proportional to their
charges.  Thus, the electric form factor of a neutral pentaquark
baryon play a role of measuring how much SU(3) symmetry is broken.
As shown in the right panel of Fig.~\ref{fig:3}, the neutral electric
form factors are negative, which leads to the positive electric charge
radii (see Table~\ref{table:1}).  We also find that the electric
form factors of the neutral pentaquark baryons have almost the same
$Q^2$ dependence.

\begin{figure}[h]
\begin{center}
\includegraphics[scale=0.57]{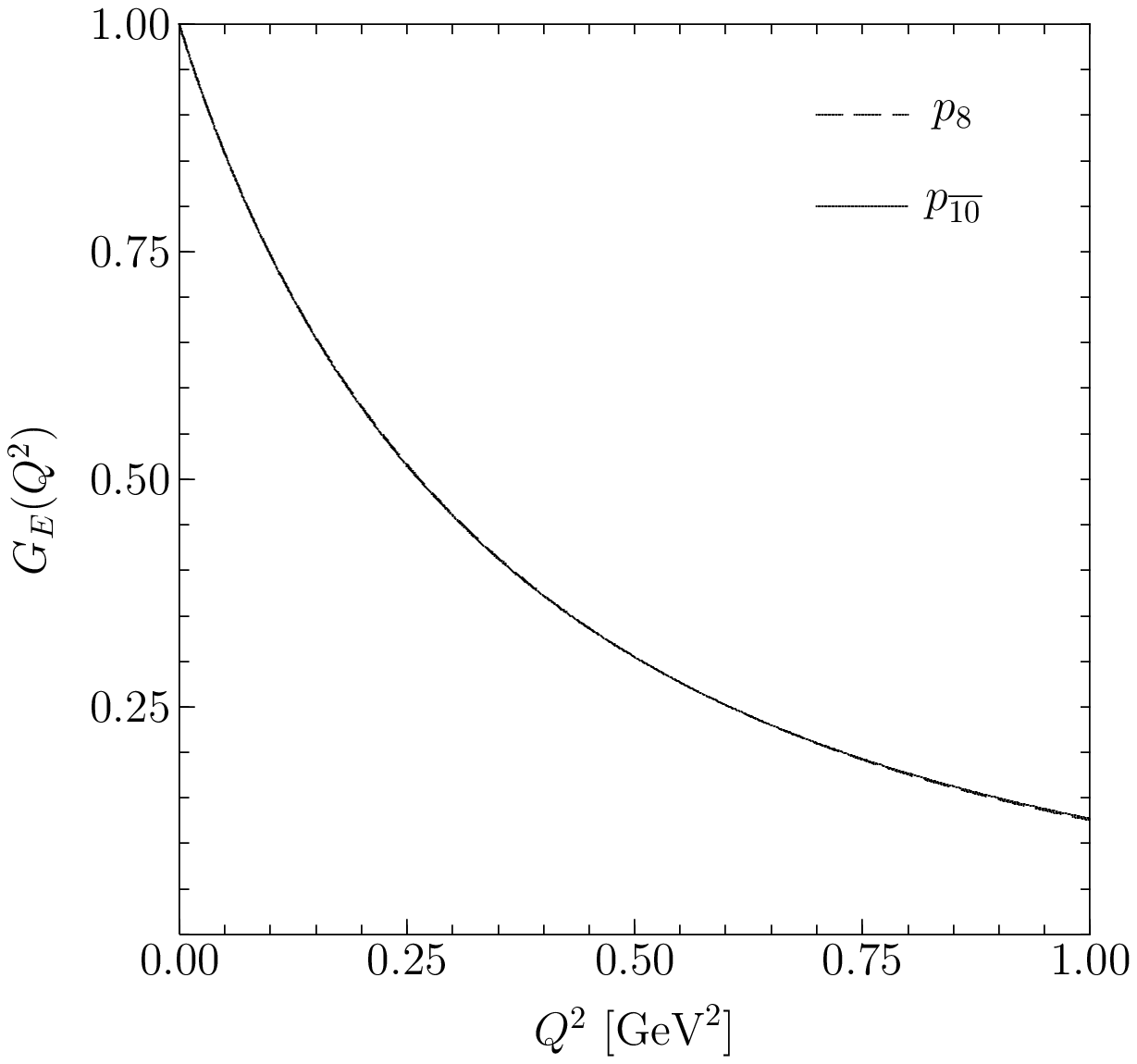} \hspace{0.4cm}
\includegraphics[scale=0.57]{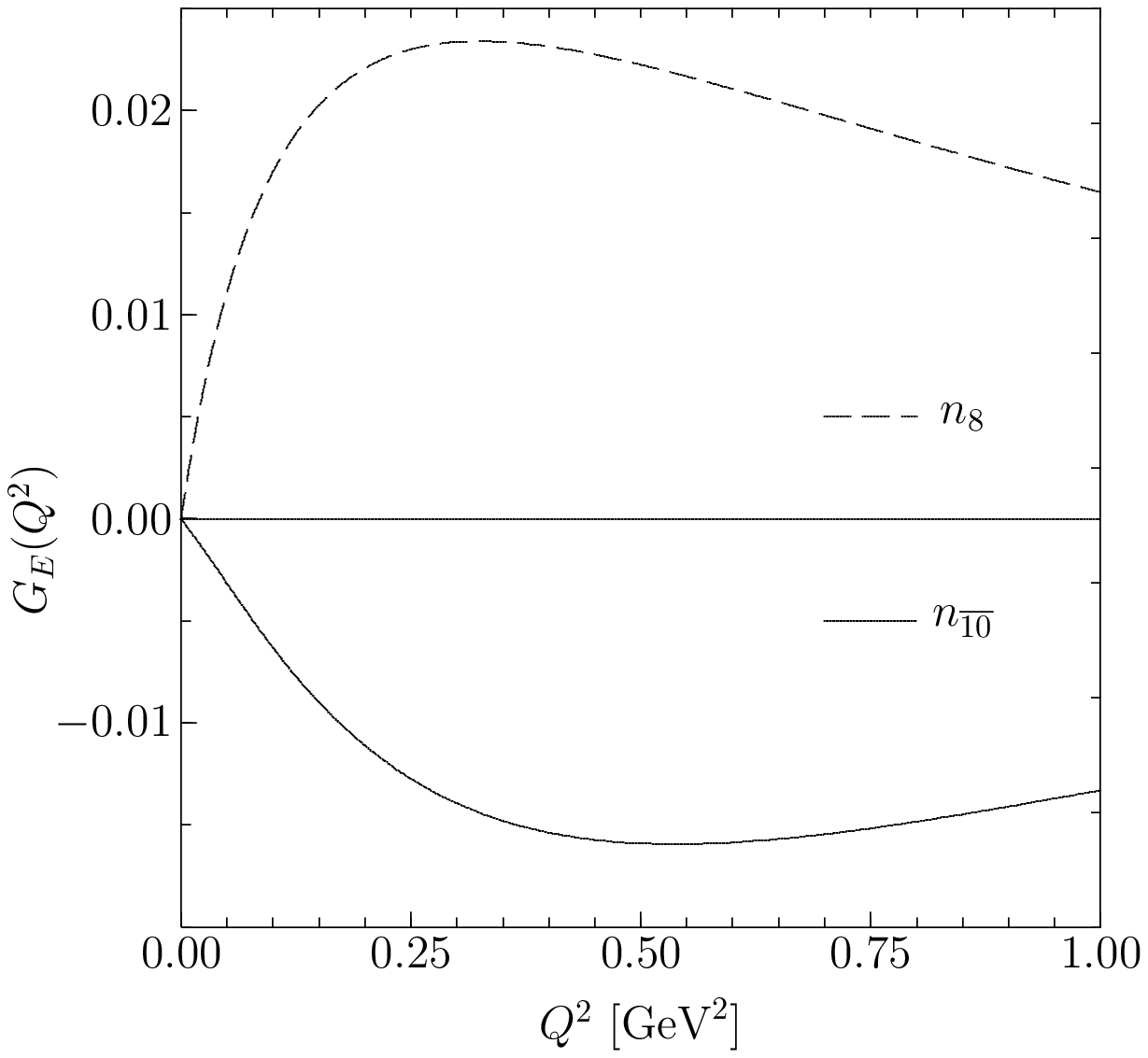}
\end{center}
\caption{Electric form factors of the nucleon and pentaquark
nucleon as functions of $Q^2$.  In the left (right) panel we
compare the proton (neutron) form factor with that of the
pentaquark proton (neutron).} \label{fig:4}
\end{figure}

Figure~\ref{fig:4} compares the electric form factors of the nucleon
with those of the pentaquark nucleon.  The electric form factor of
the pentaquark proton turns out to be almost the same as that of the
proton.  It is a remarkable result, reminding the fact that the
pentaquark proton may contain the valence $s$ and $\bar{s}$ quarks
in addition to the valence $uud$ component.  We can infer
from this result that the $s$ and $\bar{s}$ overlap each other, so
that they do not affect much the structure of the pentaquark proton.
As a result, its general structure remains very similar to that of the
proton.

In the right panel of Fig.~\ref{fig:4}, we compare the electric
form factor of the pentaquark neutron to that of the octet
neutron. As discussed above, the electric form factor of the
anti-decuplet neutron vanishes in the SU(3) symmetric case, while
that of the octet neutron does not.  Though the general shapes of
the electric form factors look similar, there is a large
difference in the vicinity of $Q^2=0$. The neutron electric form
factor changes sensitively near the zero momentum transfer as
$Q^2$ increases, whereas that of the pentaquark neutron does
slowly.  It leads to the fact that the electric charge radius of
the pentaquark neutron turns out to be very much smaller than that
of the octet neutron.

\begin{figure}[h]
\begin{center}
\includegraphics[scale=0.7]{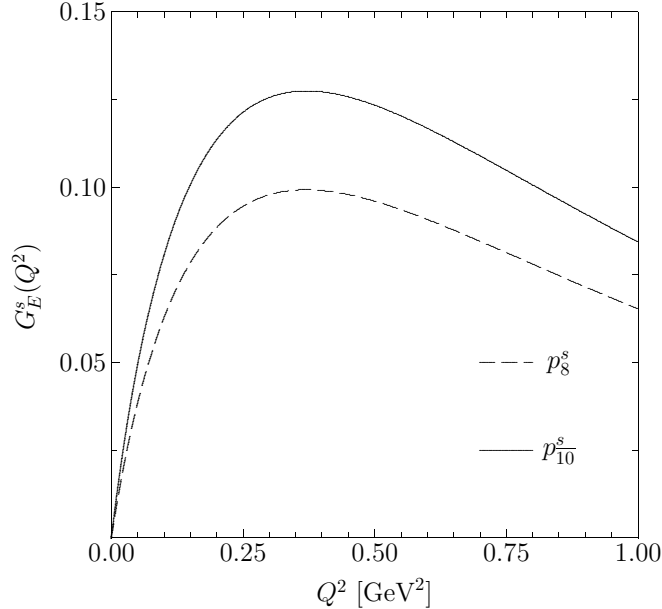}
\end{center}
\caption{Strange electric form factors of the proton and
pentaquark proton as functions of $Q^2$.  The solid curve depicts
that of the pentaquark proton, while the dashed one does that of
the proton.} \label{fig:5}
\end{figure}

Figure~\ref{fig:5} compares the strange electric form factor of
the pentaquark proton with that of the octet proton.  Since in
some models the pentaquark proton contains the valence $s$ and
$\bar{s}$ quarks, it may be expected that it has a larger strange
component than the octet proton in which the strange quark
contributions are bound to arises from the $s\bar{s}$ sea.  In this
context it is interesting that the strange electric form factor of
the octet proton has been measured very
recently~\cite{Aniol:2005zf}. As drawn in Fig.~\ref{fig:5}, the
strange electric form factor of the pentaquark proton turns out to
be larger than that of the proton by around $20\,\%$, while its
general shape is very similar to that of the proton. This is again
caused by the fact that in the $\chi$QSM the strange component is
also a sea-quark excitation and not caused by a valence quark.

{\bf 4.} In the present work, we have applied the SU(3) chiral
quark-soliton model ($\chi$QSM) to the calculation of electric
form factors of the baryon anti-decuplet particles and their
electric radii, emphasizing their difference from those of the
baryon octet. We have assumed isospin asymmetry and have
incorporated the symmetry-conserving quantization.  The rotational
$1/N_c$ and strange quark mass $m_{\rm s}$ corrections were taken
into account.  The dependence on the only free parameter $M$
turned out very mild in general and we chose the proved value
$M=420$ MeV for which many properties of the octet and decuplet
baryons are for many years known to be well reproduced.

We first have studied the flavor-decomposed electric density
distributions of the pentaquark baryon $\Theta^+$, which provide
information on the quark configuration inside the $\Theta^+$.  We
have found that the peak of the strange quark distribution is
located $0.25$ fm larger $r$, compared to that of the up and down
quarks. It implies that the strange anti-quark is with higher
probability found in the outer region of the $\Theta^+$, however
still not too far from the up and down quarks.

We have calculated the electric charge radii of the baryon
anti-decuplet.  The results turned out that those of the charged
anti-decuplet baryons are only slightly larger than those of the
charged octet baryons. It means that the pentaquark baryons are
quite compact objects, even though their leading Fock-component
corresponds to five quarks. Apparently the higher Fock-components
have a noticeable influence. Moreover, we found that the $m_s$
corrections are almost negligible to the electric charge radii.
This again indicates that the strange quarks arise from
excitations of the Dirac sea rather than being valence objects.
The electric charge radii of the neutral baryon anti-decuplet
become much smaller than those of the neutral baryon octet. The
reason can be found from the fact that the $m_s$ corrections turn
to be the leading-order contributions to them, i.e. when the
$m_{\rm s}$ corrections are switched off, the electric charge
radii of the neutral pentaquark baryons disappear. For comparison,
we also presented the electric charge radii of the baryon
decuplet, which are found to be larger than those of the baryon
octet and anti-decuplet.

We also have examined the electric form factors of the pentaquark
baryons.  Since the electric form factors of the anti-decuplet
baryons are proportional to their charges in the SU(3) symmetric
case, their general tendency is simply the same in SU(3) symmetry.
Since the terms of the $m_s$ corrections are destructively
interfering, the SU(3) symmetry-breaking effects do not play any
significant role in these form factors. However, for the neutral
pentaquark baryons the $m_s$ corrections turn out to provide the
main contributions.  Hence, these can be regarded as a measure of
the degree of SU(3) symmetry breaking.

We have compared the electric form factors of the pentaquark
nucleon with those of the nucleon.  The electric form factor of
the pentaquark proton are, remarkably, almost the same as that of
the proton.  It implies that the valence $s$ and $\bar{s}$ quarks
in the pentaquark proton overlap each other, so that they do not
influence much the structure of the pentaquark proton.  It means
again that the pentaquark proton is a compact baryon despite of
its five-quark structure.  As for the neutron electric form
factors, their general shapes look similar.  However, it is found
that there is a large difference in the vicinity of $Q^2=0$ where
the form factor of the octet neutron changes much faster than the
one of the anti-decuplet neutron.

We finally have investigated the strange electric form factors of the
pentaquark proton.  The pentaquark proton may contain the valence $s$
and $\bar{s}$ quarks, whereas the strange quark component of the
proton arises from the $s\bar{s}$ sea.  Thus, we anticipate that the
pentaquark proton might have a larger strange component than that of
the proton. Indeed, the strange electric form factor of the
pentaquark proton turns out to be larger than that of the proton by
around $20\,\%$.

In conclusion, we have found two significant features of the
pentaquark baryons by studying their electric properties: Firstly,
even though their leading Fock-component consists of five quarks,
they are rather compact baryons, being comparable to the baryon
octet. Secondly, the effects of SU(3) symmetry breaking do not
play any important role in the case of the charged pentaquark
baryons. However, those effects are leading-order contributions
for the neutral ones.  As a result, electric properties of the
neutral pentaquark baryons show the extent of SU(3) symmetry
breaking. We plan further studies on other properties of the
pentaquark baryons such as magnetic transition and axial-vector
form factors, which will appear soon elsewhere.

\section*{Acknowledgments}
The authors are very grateful to M. Prasza\l owicz for helpful
discussions and comments. The present work is supported by a
Korean-German (F01-2004-000-00102-0) grant of the KOSEF and of the
Deutsche Forschungsgemeinschaft (DFG). The work of HChK is also
supported by the Korea Research Foundation Grant funded by the
Korean Government(MOEHRD) (KRF-2005-202-C00102).  The work is
partly supported by the Transregio-Sonderforschungsbereich
Bonn-Bochum-Giessen, the Verbundforschung of the Federal Ministry
for Education and Research (BMBF) of Germany, the
Graduiertenkolleg Bochum-Dortmund, the COSY-project J\"ulich as
well as the EU Integrated Infrastructure Initiative Hadron Physics
Project under contract number RII3-CT-2004-506078.

\end{document}